\UseRawInputEncoding 
\documentclass[footinbib,aip,reprint]{revtex4-1}

\usepackage{graphicx}
\usepackage{amsmath}
\usepackage{mathtools}
\usepackage{amssymb}

\usepackage{graphicx}
\usepackage{bm}

\begin{document}

\title{Magnetostriction of $\alpha$-RuCl$_3$ flakes in the zigzag phase}

\author{Yun-Yi Pai}
\email{yunyip@ornl.gov}
\address{Materials Science and Technology Division, Oak Ridge National Laboratory, Oak Ridge, TN 37831}
\address{Quantum Science Center, Oak Ridge, Tennessee 37831}

\author{Claire E. Marvinney}
\address{Materials Science and Technology Division, Oak Ridge National Laboratory, Oak Ridge, TN 37831}
\address{Quantum Science Center, Oak Ridge, Tennessee 37831}

\author{Matthew A. Feldman}
\address{Materials Science and Technology Division, Oak Ridge National Laboratory, Oak Ridge, TN 37831}
\address{Department of Physics and Astronomy, Vanderbilt University, Nashville, TN 37203}

\author{Brian Lerner}
\address{Materials Science and Technology Division, Oak Ridge National Laboratory, Oak Ridge, TN 37831}

\author{Yoong Sheng Phang}
\address{Department of Physics and Astronomy, University of Georgia, Athens, GA 30602}

\author{Kai Xiao}
\address{Center for Nanophase Materials Sciences, Oak Ridge National Laboratory, Oak Ridge, TN 37831}

\author{Jiaqiang Yan}
\address{Materials Science and Technology Division, Oak Ridge National Laboratory, Oak Ridge, TN 37831}
\address{Quantum Science Center, Oak Ridge, Tennessee 37831}

\author{Liangbo Liang}
\address{Center for Nanophase Materials Sciences, Oak Ridge National Laboratory, Oak Ridge, TN 37831}

\author{Matthew Brahlek}
\address{Materials Science and Technology Division, Oak Ridge National Laboratory, Oak Ridge, TN 37831}

\author{Benjamin J. Lawrie}
\email{lawriebj@ornl.gov; This manuscript has been authored by UT-Battelle, LLC, under contract DE-AC05-00OR22725 with the US Department of Energy (DOE). The US government retains and the publisher, by accepting the article for publication, acknowledges that the US government retains a nonexclusive, paid-up, irrevocable, worldwide license to publish or reproduce the published form of this manuscript, or allow others to do so, for US government purposes. DOE will provide public access to these results of federally sponsored research in accordance with the DOE Public Access Plan (http://energy.gov/downloads/doe-public-access-plan). }
\address{Materials Science and Technology Division, Oak Ridge National Laboratory, Oak Ridge, TN 37831}
\address{Quantum Science Center, Oak Ridge, Tennessee 37831}

\date{\today}

\begin{abstract}
Motivated by the possibility of an intermediate U(1) quantum spin liquid phase in out-of-plane magnetic fields and enhanced magnetic fluctuations in exfoliated $\alpha$-RuCl$_3$ flakes, we study magneto-Raman spectra of exfoliated multilayer $\alpha$-RuCl$_3$ in out-of-plane magnetic fields  of -6 T to 6 T at temperatures of 670 mK - 4 K. While the literature currently suggests that bulk $\alpha$-RuCl$_3$ is in an antiferromagnetic zigzag phase with R$\overline{3}$ symmetry at low temperature, we do not observe R$\overline{3}$ symmetry in exfoliated $\alpha$-RuCl$_3$ at low temperatures. While we saw no magnetic field driven transitions, the Raman modes exhibit unexpected stochastic shifts in response to applied magnetic field that are above the uncertainties inferred from Bayesian analysis.  These stochastic shifts are consistent with the emergence of magnetostrictive interactions in exfoliated $\alpha$-RuCl$_3$.

\end{abstract}


\maketitle

\section{Introduction}
The quantum spin liquid (QSL) is a long sought-after non-classical phase characterized by a topological order parameter~\cite{ANDERSON1973153, Knolle2019, Savary2016}.  QSLs may be critical to the development of topologically protected quantum computing platforms because they may host non-local excitations with Anyonic statistics.  Amongst possible candidates that could host such a topologically protected phase, $\alpha$-RuCl$_3$ has been extensively studied over the past several years because it can be approximately described by the analytically solvable Kitaev honeycomb model\cite{KITAEV20062}.  Several experimental efforts have reported features consistent with a QSL phase, including half-quantized thermal Hall conductance plateaus \cite{Kasahara2018}, a scattering continuum in Raman spectroscopy \cite{Sandilands2015, WangYiping2020, Wulferding2020}, neutron scattering \cite{Banerjee2016NM, Banerjee1055}, and nuclear magnetic resonance \cite{Baek2017, Jansa2018}. 

Despite numerous reports of possible QSL signatures, many fundamental questions remained unanswered, both theoretically and experimentally.  For example, while the room temperature structure is now accepted as C2/m\cite{Lebert_2020, Mai2019}, the low temperature symmetry of $\alpha$-RuCl$_3$ remains in question.  Early reports suggested trigonal P3$_1$12 \cite{Monoclinic_Johnson} and monoclinic C2/m \cite{Monoclinic_Johnson, C3mCao, STMc3m} symmetries, but more recent reports have suggested the presence of rhombohedral R$\overline{3}$ \cite{2016arXiv160905690P, GlamazdaHytersis} symmetry.  Further, it remains unclear whether the Kitaev model exchange parameters should be antiferromagnetic or ferromagnetic.  Additionally, a recent density functional renormalization group (DMRG) calculation suggested that commonly reported QSL phases induced by in-plane magnetic fields are missing in the DMRG result, but a U(1) QSL phase can be stabilized by out-of-plane magnetic fields \cite{Jiang_2019}. The number of antiferromagnetic zigzag phases that exist before the onset of the QSL phase also remains in question, as does the effect of sample-to-sample variability\cite{kim2021investigation} between ostensibly identical flakes and between exfoliated flakes and bulk crystals.   

The layered Van der Waals nature of $\alpha$-RuCl$_3$ enables the possibility of heterostructure assembly, proximity effect engineering, and strain engineering \cite{Geim_Review, An_Review_2D, Universal,GrapheneRuCl, ModulationDope}. Exfoliated flakes a few layers thick are air-stable\cite{FlakeEnhanced, notstable}, greatly increasing the flexibility of the workflow.  On the other hand, reduced dimensionality yields stronger order-parameter fluctuations and eventually suppression of long-range order\cite{MerminWagner}.  Stacking faults may also open up additional hopping pathways to stabilize the QSL phase\cite{FlakeEnhanced}. Magnetic fluctuations have been reported to persist \cite{Lin2020} or be enhanced \cite{FlakeEnhanced, notstable} in the few-layer to monolayer limit.  Furthermore, strain gradients may induce synthetic gauge fields that locally tune topological phases\cite{GaugeField1, GaugeField2}.  A recent first principle calculation suggests that 2\% strain is enough to drive monolayer $\alpha$-RuCl$_3$ from the AFM zigzag phase to the spin-polarized phase. Topological devices with well-patterned QSL puddles may be possible with appropriate gauge field engineering. 

Raman spectroscopy is a flexible and powerful tool for resolving sample symmetry as well as microscopic electron-phonon and phonon-magnon interactions. In addition to the above described characterization of C2/m and R$\overline{3}$ symmetry, Raman spectroscopy has been used to characterize temperature dependent hysteresis \cite{GlamazdaHytersis}, several magnon modes, and a possible Majorana mode \cite{Wulferding2020, HighfieldSahasrabudhe}.  Here, motivated by a possible intermediate U(1) QSL phase in out-of-plane magnetic fields \cite{Jiang_2019} and larger tunability in exfoliated $\alpha$-RuCl$_3$, we study the Raman spectra of exfoliated $\alpha$-RuCl$_3$ in out-of-plane magnetic fields in a Faraday geometry\cite{Wulferding2020} at temperatures as low as $T = $ 670 mK. 

\section{Sample Details and Experimental Setup }

\begin{figure}[t]
\centering
    \includegraphics[width=\columnwidth]{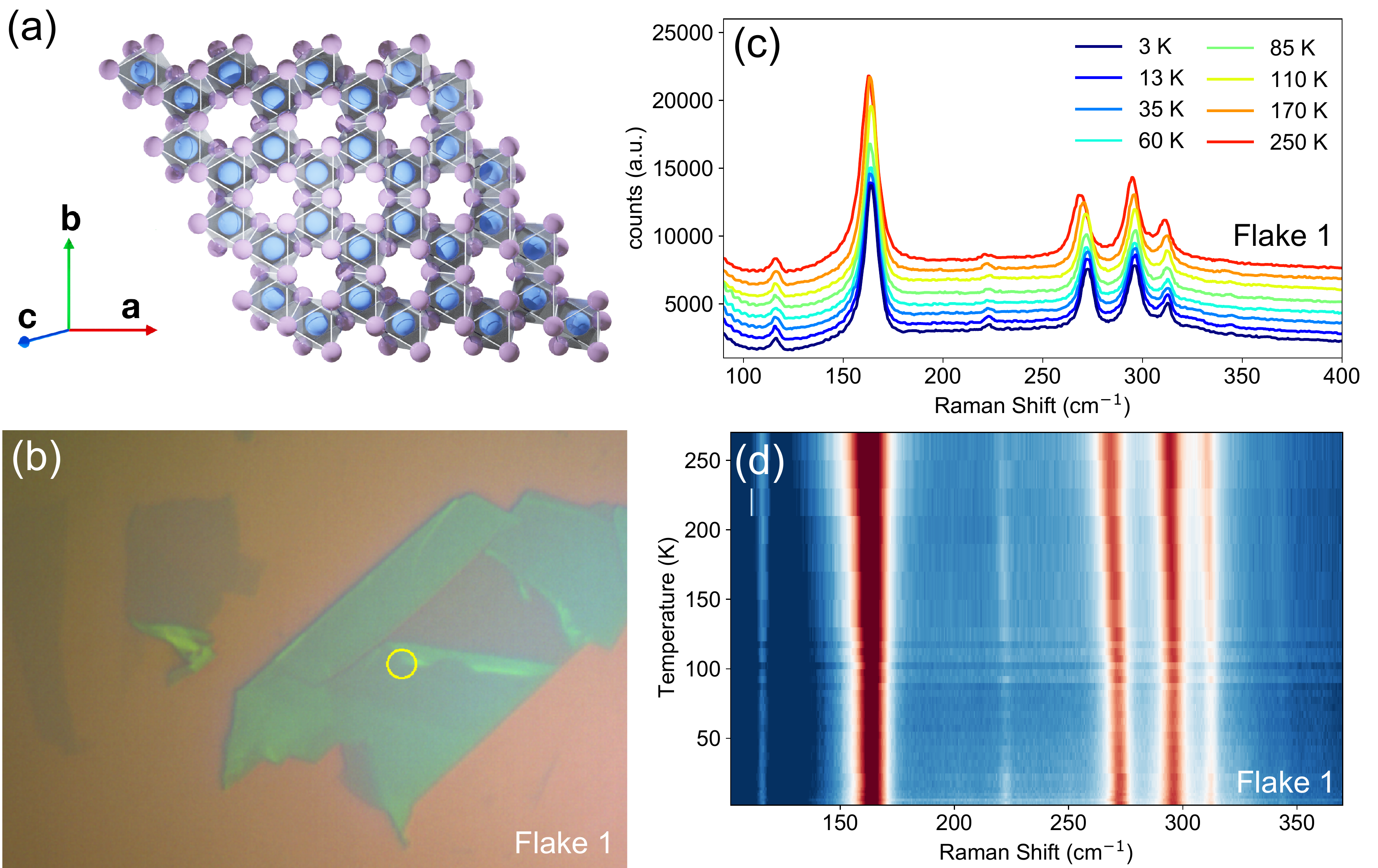}
    \caption{(a) $\alpha$-RuCl$_3$ lattice structure. (b) optical image of flake 1. (c) Raman spectra of flake 1 from $T = $ 3 K to 300 K. (d) intensity plot of the Raman spectra in (c). 
}
    \label{fig:crystal_and_temp}
\end{figure}

Bulk $\alpha$-RuCl$_3$ single crystals were grown by vapor transport \cite{may2020practical}. The $\alpha$-RuCl$_3$ flakes were mechanically exfoliated onto a 300 nm SiO$_2$ film on a Si substrate.  Due to signal-to-noise-ratio (SNR) constraints and large parameter sweeps, we focused on thick flakes with lateral dimensions of tens of microns. Variable temperature Raman spectra were acquired for temperatures of 3 K - 300 K in a Montana Instruments closed-cycle cryostat with a Princeton Instruments Isoplane SCT-320 spectrograph, a Pixis 400BR Excelon camera, and a 2400 line/mm grating.  The sample was probed in a backscattering configuration (beam path $\|$ \textbf{c*}, where \textbf{c*} is the outer product of the crystallographic \textbf{a} and \textbf{b} axes illustrated in Figure \ref{fig:crystal_and_temp} (a)) with a 532.03 nm excitation at 2.0 mW power and 45 sec integration time.  The laser excitation was removed with Semrock 532 nm RazorEdge ultrasteep dichroic and long-pass edge filters with cutoff at 90 cm$^{-1}$. 

The $T = $ 4 K and 670 mK magnetic field dependent Raman spectra were acquired in a customized Leiden dilution refrigerator with free space optics access that is described in further detail elsewhere\cite{lawrie2021freespace}.  Raman spectra were recorded with an Andor Kymera 193 spectrograph with a Newton EMCCD DU970P-BV camera and a 2400 line/mm grating.  The exposure time was 1800 seconds per spectrum. The sample was probed in a backscattering configuration with a 2 mW (200 uW), 532.2096 nm laser excitation at 4K (670 mK). The excitation was filtered by a set of 3 volume Bragg gratings (Optigrate, 1 volume Bragg beam splitter and 2 volume Bragg notch filters). The Faraday geometry (B $\|$ \textbf{c*} and beam path) induces a $\theta_F (B) = -25.60^{\circ}$/T polarization rotation due to the beam propagation through the objective.   

Initial temperature dependent Raman spectra acquired as flake 1 (illustrated in Figure \ref{fig:crystal_and_temp} (b)) warmed from $T = $ 3 K to 270 K  are shown in Figure \ref{fig:crystal_and_temp} (c) and (d).  All the Raman modes here are consistent with those reported in the literature.  Using peak assignments from the R$\overline{3}$ space group\cite{Gaomin2019} for low temperature, we identify E$^1_g$ at 116 cm$^{-1}$, E$^2_g$ at 164 cm$^{-1}$, E$^3_g$ at 272 cm$^{-1}$, E$^4_g$ at 296 cm$^{-1}$, and A$^1_1g$ at 313 cm$^{-1}$. We note that the low energy tail below 125 cm$^{-1}$ is not negligible and results in distortion of the lineshape of the E$^1_g$ Fano peak at 116 cm$^{-1}$ in this dataset (this is not the case for the other datasets taken with 3 volume Bragg gratings). 

Additionally, less reported $\alpha$-RuCl$_3$ Raman modes at 222 cm$^{-1}$ and 345 cm$^{-1}$ are observed. The 222 cm$^{-1}$ peak has been attributed to stacking faults\cite{Lin2020} or defects\cite{Gaomin2019}.  It has been reported that thin flakes are more prone to stacking faults than single crystals\cite{kim2021investigation}, which is consistent with the fact that the $\alpha$-RuCl$_3$ flakes studied here are in the thin crystal to thick flake limit. The 345 cm$^{-1}$ peak has been attributed to A$^2_{1g}$ due to its $XX$ polarization (parallel polarization) nature\cite{Lin2020} and to defects\cite{Gaomin2019}. Notably, Li et al. \cite{Gaomin2019} reported that both the 222 cm$^{-1}$ peak and 345 cm$^{-1}$ peak only showed up for blue (488 nm) excitation but not for red (633 nm) excitation. 

\begin{figure*}[t]
\centering
    \includegraphics[width=2\columnwidth]{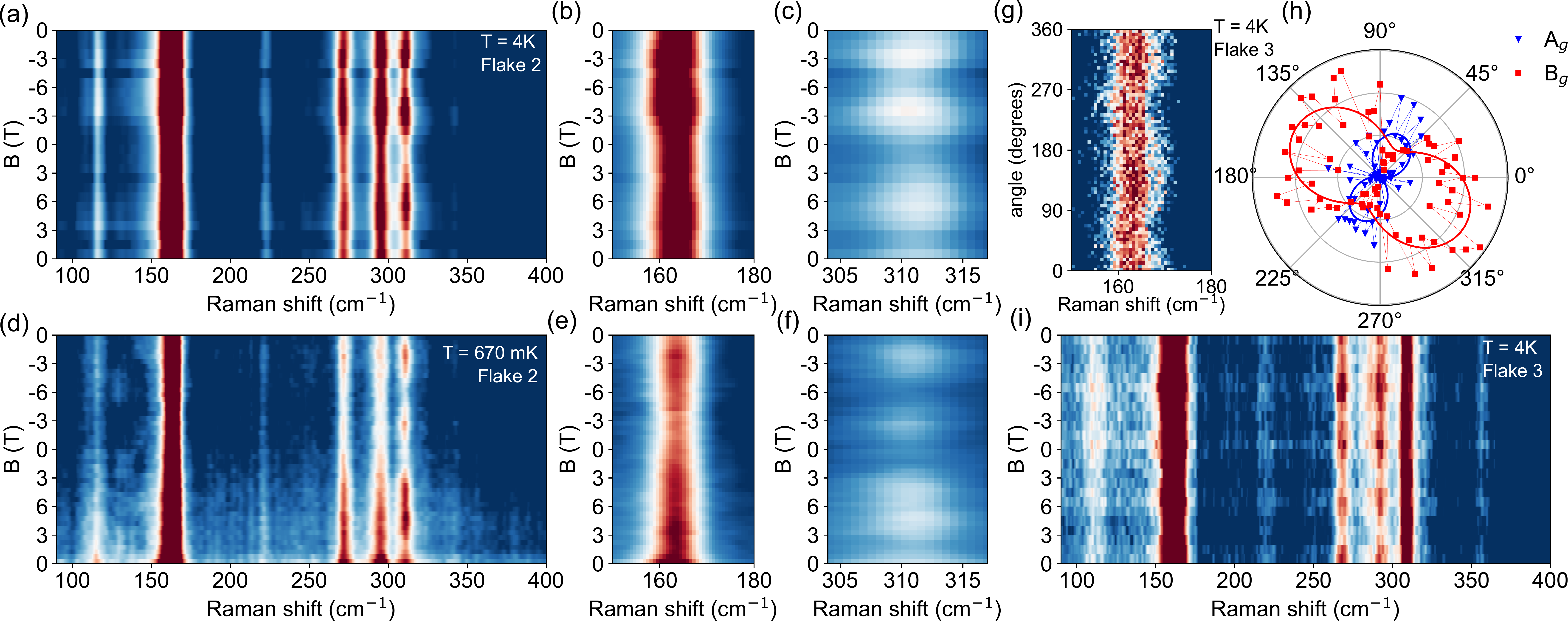}
    \caption{Raman spectra as a function of out-of-plane magnetic field at (a) $T = $ 4 K and (d) $T \sim $ 670 mK. (b) The 164 cm$^{-1}$ and (c) the 313 cm$^{-1}$ Raman modes exhibit a clear fluctuation in intensity and energy with magnetic field. At 670 mK, the 164 cm$^{-1}$ (e) and 313 cm$^{-1}$ (f) modes also exhibit a magnetic field dependence. The signal-to-noise ratio for the spectra at $T = $ 670 mK is reduced by $\sim \sqrt{10}$ because of the reduced laser power and unchanged integration time. (g) angular dependence of the 164 cm$^{-1}$ mode at $B = 0$ T with $XX$ (parallel) polarization configuration. (h) Intensity of the A$_g$ and B$_g$ modes as a function of the angle $\alpha$ between polarization and the sample orientation from (g). (i) Raman spectra as a function of field with $XX$ (parallel) polarization configuration and a constant $\alpha$ at $T = 4$ K.}
    \label{fig:4KMK}
\end{figure*}

\section{Magneto-Raman spectroscopy}

Figure \ref{fig:4KMK} (a) shows the measured Raman spectra of flake 2 as a function of out-of-plane magnetic field ($B$ $\|$ $\textbf{c}^*$) at $T = $ 4 K with the magnetic field swept from $B = 0$ T, $+$6 T, $-$6 T, to $0$ T in 1 T increments. Figure \ref{fig:4KMK} (d) shows the results of the same measurement repeated at $T \sim $ 670 mK with 0.5 T steps. Two clear observations can be made: 
\begin{enumerate}
  \item  The absolute intensity of the 313 cm$^{-1}$ peak has a clear asymmetric behavior as a function of magnetic field; it has local minima at  $B \sim -$ 6 T and $B \sim +$ 1 T and a local maximum at $B \sim +$ 5 T.
   \item Almost all the peaks appear to fluctuate as a function of magnetic field, an effect that is clear in the magnified 164 cm$^{-1}$ and 313 cm$^{-1}$ modes shown in Figure \ref{fig:4KMK} (b-c) and (e-f). 
\end{enumerate}

The first observation can be simply explained by the fact that the angle $\alpha$ between the polarization and the crystal \textbf{a}, \textbf{b} axes was not fixed during the magnetic field sweep. The angle rotates at $\theta_F (B) = -25.60^{\circ}$/T due to Faraday rotation in the objective. Since the 313 cm$^{-1}$ mode is sensitive to XX polarization only, it is known to have a cos(2$\alpha$) dependence\cite{Mai2019}, where $\alpha$ is the angle between the polarization and \textbf{a} axis. The observed magnetic-field dependence of the 313 cm$^{-1}$ peak intensity is consistent with this polarization-rotation effect. 
The same explanation could, at first glance, be true for the second observation: as reported in Mai et al. \cite{Mai2019}, the spectral weights of the modes from the $A_g$ series and $B_g$ series are a function of $\alpha$. It is natural to consider the possibility that a similar effect may be present here. However, the measurement reported by Mai et al. \cite{Mai2019} was performed at room temperature, and the irreducible representation of the space group C2/m was assumed. For the low temperature spectra reported here, the likely space group R$\overline{3}$ yields phonon modes at similar energies, and the decomposed $E_g$ modes, though doubly degenerate, are not broken. Fluctuations as a function of magnetic field have not previously been observed for low temperature polarization resolved Raman spectroscopy of single crystal $\alpha$-RuCl$_3$\cite{Priv_Comm_Thuc}. 

We subsequently performed a polarization sweep at zero magnetic field on flake 3. Figure \ref{fig:4KMK} (g) illustrates the normalized 164 cm$^{-1}$ mode as a function of the polarization angle. A clear 2-fold oscillation is observed. Fitting the 164 cm$^{-1}$ peak to a pair of closely spaced peaks yields Figure \ref{fig:4KMK} (h).  This angular dependence is different from the expected R$\overline{3}$ angular independence \cite{Priv_Comm_Thuc}. It is also different from the reported angular dependence for C2/m, which has a 4-fold symmetry\cite{Mai2019}. However, it is worth pointing out that fits of the peak position may be affected by the baseline, which could, in principle, be affected by the angle $\alpha$.

Figure \ref{fig:4KMK} (i) illustrates a set of Raman spectra on flake 3 as a function of magnetic field at $T = 4$ K with parallel $XX$ polarization configuration and a constant polarization angle $\alpha$ with respect to the flake. This was done by compensating half of the $\theta_F (B) = -25.60^{\circ}$/T on both the excitation and collection path. We note that the 345 cm$^{-1}$ peak on this flake appears at 359 cm$^{-1}$, consistent with previous reports that this peak is related to stacking faults or defects \cite{Gaomin2019}) that may vary significantly from flake to flake.  

\begin{figure}[t]
\centering
    \includegraphics[width=\columnwidth]{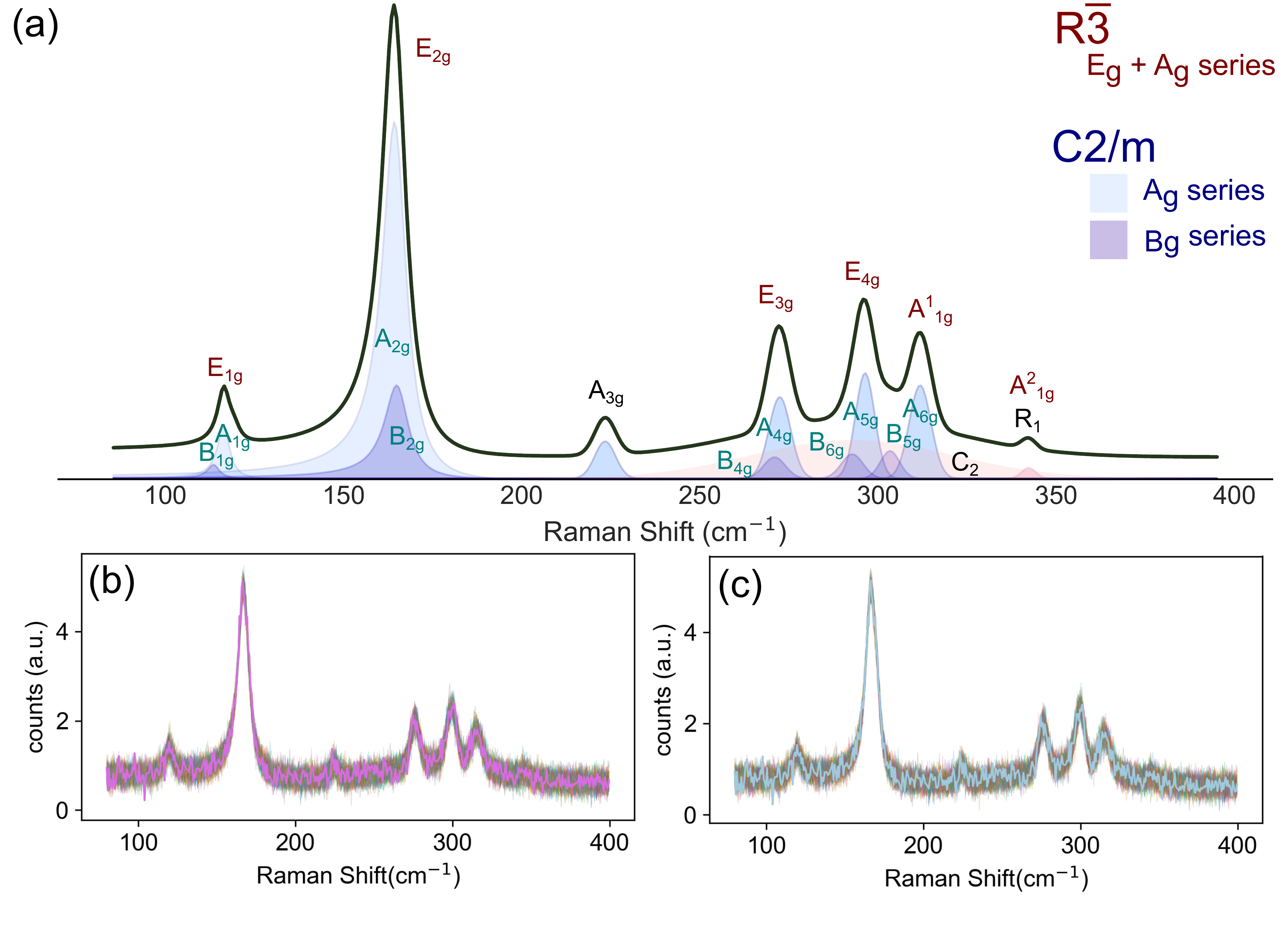}
    \caption{(a) Peak assignment from R$\overline{3}$ and C2/m space groups. The R$\overline{3}$ has doubly degenerate $E_{g}$s and singlet $A_{g}$s. The peaks in C2/m consist of $A_g$ series and $B_g$ series. (the solid line is the resulting spectrum.) (b) An example spectrum taken at $T = 670$ mK fit with R$\overline{3}$ peaks. (c) The same example spectrum taken at $T = 670$ mK fit with C2/m peaks. 
}
    \label{fig:fit}
\end{figure}

\begin{figure*}[t]
\centering
    \includegraphics[width=2\columnwidth]{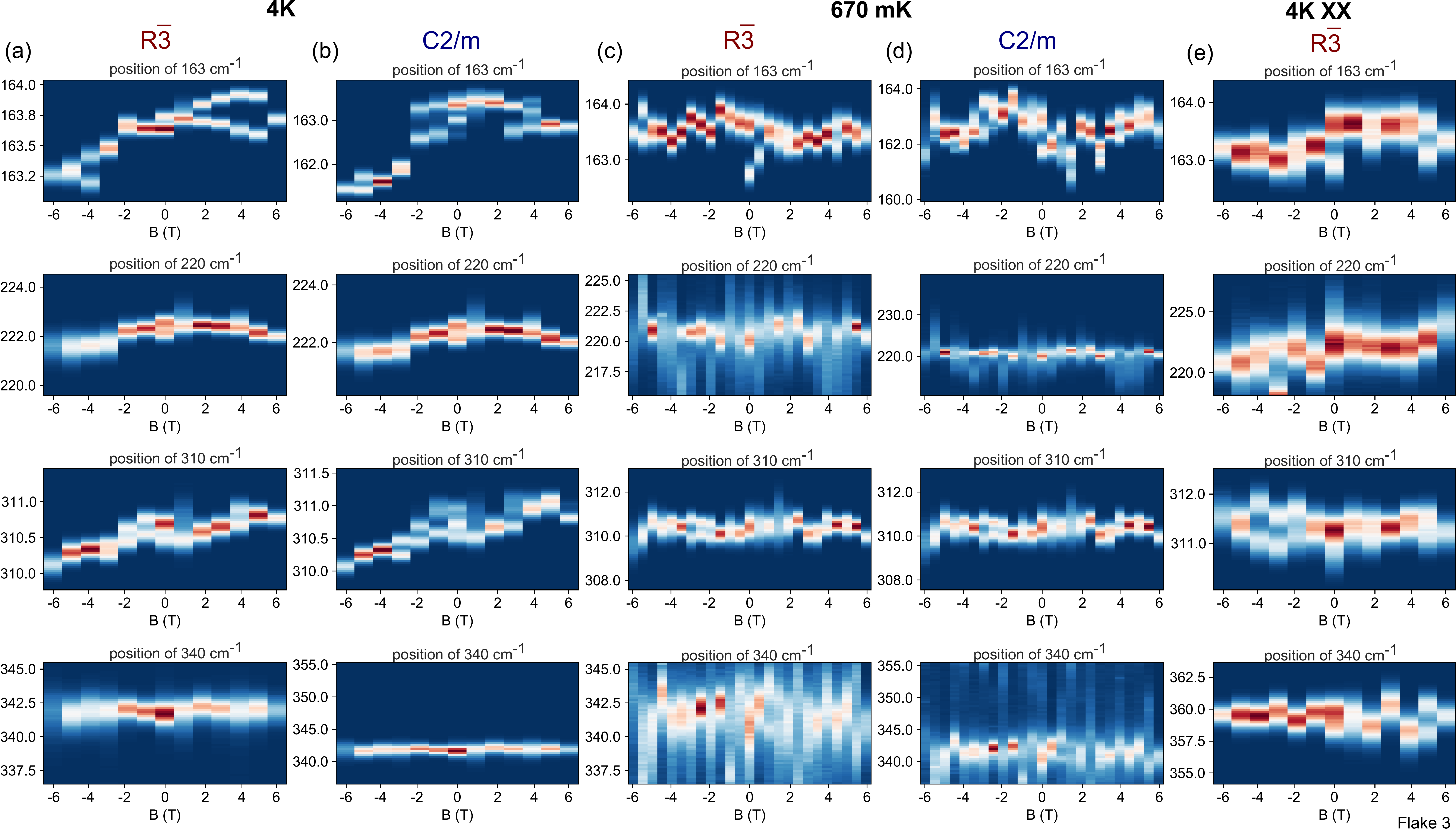}
    \caption{Posterior distributions from Bayesian inference. (a) Selected inferred peak positions using R$\overline{3}$ peak assignment for spectra taken at $T = 4$ K. (b) Selected inferred peak positions using C2/m peak assignment for spectra taken at $T = 4$ K. (c) Selected inferred peak positions using R$\overline{3}$ peak assignment for spectra taken at $T = 670$ mK. (d) Selected inferred peak positions using C2/m peak assignment for spectra taken at $T = 670$ mK. (e) Selected inferred peak positions using R$\overline{3}$ peak assignment for spectra taken at $T = 4$ K with parallel ($XX$) polarization. 
}
    \label{fig:bayes}
\end{figure*}

\section{Bayesian Inference}
In order to better understand the Raman spectra shown above, we employed Bayesian inference techniques, considering both the modes from R$\overline{3}$ with unsplit degeneracies and C2/m. For R$\overline{3}$, as shown in Figure \ref{fig:fit} (a), the model consists of 2 Fano lineshapes ($E_{1g}$, $E_{2g}$), 5 Gaussian peaks ($E_{3g}$, $E_{4g}$, $A^1_{1g}$, $A_{3g}$, and $R_1$). To capture the baseline, we use a linear background with a broad Gaussian centered at $\sim$ 300 cm$^{-1}$. For C2/m, peaks consist of the $A_g$ series and $B_g$ series. The A$_{1g}$, B$_{1g}$, A$_{2g}$, B$_{2g}$ are fitted to Fano lineshapes and the rest to Gaussian. We fit the Raman spectra from 90 cm$^{-1}$ to 400 cm$^{-1}$ in a single step with \texttt{Scipy}. We then used the fitting result as the starting center of the distribution of our priors. The priors of the parameters are assumed to be Gaussian. The Bayesian inference was done using Hamiltonian Monte Carlo \texttt{PyMC3}\cite{pyMC}. For fast convergence we use a No U-Turns (NUTS) sampler. We use 4 chains with 3,000 samples per chain. It takes between 3 and 100 minutes for a spectrum to converge, depending on the complexity of the models.

\begin{figure}[t]
\centering
    \includegraphics[width=\columnwidth]{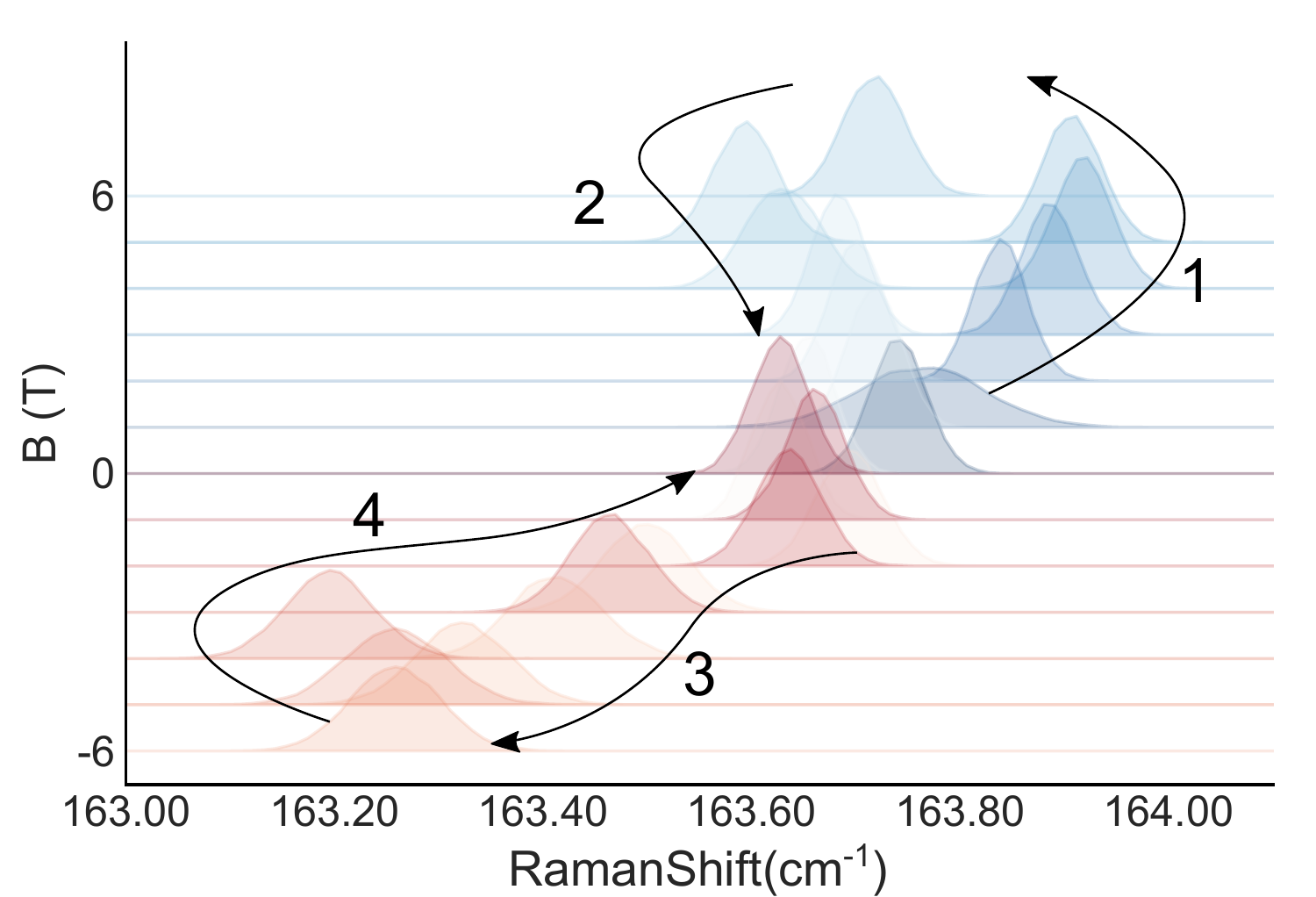}
    \caption{Hysteretic frequency shift of the 163 cm$^{-1}$ mode. Waterfall plot of the posterior distribution of the 163 cm$^{-1}$ mode position as a function of the magnetic field (from Fig. \ref{fig:bayes}) (a). As the magnetic field is swept from 0 T $\to$ $+6$ T $\to$ 0 T $\to$ $-$ 6 T $\to$ 0 T, the peak position follows the arrow 1 $\to$ 2 $\to$ 3 $\to$ 4. This hysteretic response appears in Bayesian inference as well as frequentist fits. The hysteresis is statistically significant.  For example, from the Bayesian inferred result of the first spectrum at $B$ = 0 T, the mean = 163.74 cm$^{-1}$, the 3\% HDI = 163.69 cm$^{-1}$, and the 97\% HDI = 167.79 cm$^{-1}$, and thus the middle 94\% HDI = 0.10 cm$^{-1}$ with a standard deviation = 0.027 cm$^{-1}$).  For this case, the size of the hysteresis of the loop of the positive field is about 0.19 cm$^{-1}$, which is 7.4 $\sigma$.}
    \label{fig:hysterisis}
\end{figure}

Figure \ref{fig:fit} (b) shows the resulting reconstructed spectra from the posterior predictives using R$\overline{3}$ peak assignments plotted with an original spectrum taken at $T = 670$ mK.  Figure \ref{fig:fit} (c) shows the resulting reconstructed spectra from the posterior predictives using C2/m peak assignments plotted with an original spectrum taken at $T = 670$ mK. Figure \ref{fig:bayes} shows the posterior distribution. Note that these histograms are the distributions of the posteriors at each field. Since the magnetic field was swept from $B = 0$ T to $B = +$6 T to $B = -$6 T to $B = 0$, the extrema are singly visited, $B = 0$ T is triply visited, and all the other $B$ field values are doubly visited. No manual normalization was applied. We see: 

\begin{enumerate}
  \item While the uncertainty for each posterior peak position is different, most of the peaks shown here move above their highest density interval (HDI) \cite{HDI}.
  \item While the models R$\overline{3}$ and C2/m are very different, the posterior peak positions for 222 cm$^{-1}$, 313 cm$^{-1}$, and 341 cm$^{-1}$ are quantitatively similar within a dataset. This means that these peak positions are robust against bias from the details of the other peaks. The joint scatter plot of peak position against, for example, the baseline, are essentially orthogonal. Hence the change of the peak position as a function of the magnetic field is not likely to be introduced by a polarization dependent baseline.  
  \item While polarization dependent frequency shifts exist for some of the modes such as the 163 cm$^{-1}$ mode for C2/m symmetry, the 222 cm$^{-1}$ and 313 cm$^{-1}$ modes are known to not exhibit frequency shifts as a function of the polarization angle. Furthermore, the polarization induced frequency shift has clear rotational symmetry that is missing here. 
  \item The 163 cm$^{-1}$ mode has 2 clear hysteresis loops in the R$\overline{3}$ model, as shown in Figure \ref{fig:hysterisis} that is less clear in the C2/m model. The posterior for the 163 cm$^{-1}$ mode in the C2/m model has a multinomial distribution, which is usually an indication of the model not converging well, possibly due to the complexity of the model, which includes multiple closely spaced Raman modes.   
  \item The trend is different each sweep, even for the same flake. For example, the 310 cm$^{-1}$ peak has a positive slope as a function of magnetic field in the first sweep at $T = 4$ K, but not in the second sweep at $T = 670$ mK.
\end{enumerate}

\section{Discussion and Conclusion}
The variation in vibrational Raman modes as a function of magnetic field reported here is consistent with spin-phonon interactions and magnetostrictive effects.  Recently Gass et al. \cite{Gass2020} and Schonemann et al. \cite{Schonemann2020} reported contraction along the \textbf{c$^{*}$} axis and along the in-plane direction in response to applied in-plane fields. Though not reported yet, it is natural to assume similar effects may occur in response to out-of-plane magnetic fields as well. While Raman shifts seemed to be negligible in a previous study of single crystals \cite{Wulferding2020}, magnetostrictive effects may be stronger in exfoliated flakes due to reduced rigidity.  On the other hand, related materials like CrI$_3$ have strong Raman shifts for out-of-plane magnetic fields of $B = 1$ T\cite{DistinctCrI3}.  A non-reciprocal magneto-phonon interaction was also reported for CrI$_3$ in the form of polarization rotation asymmetric with magnetic field \cite{Liueabc7628}, but little to negligible shift of the vibrational Raman modes was described in that report.  In addition to CrI$_3$, Raman frequency shifts in response to magnetic field have been observed in 15R-BaMnO$_3$\cite{MagnetoStriction15R}.

The hysteretic shifts in $\alpha$-RuCl$_3$ Raman modes reported here are statistically significant, but highly variable. This could plausibly be due to variation in microscopic spin arrangement and strain that can result from substrate roughness \cite{builtintension} or from variations in flake geometry and thickness and may also affected by the exfoliation process. For example, in CrI$_3$, Raman modes depend on the spin arrangement and are sensitive to spin flips\cite{DistinctCrI3}. As the magnetic field is applied, the spins in $\alpha$-RuCl$_3$ flakes may re-orient, and the flake itself may deform mechanically. As the magnetic field is removed, the flake may settle in a new spin-mechanical configuration that is more stable than the initial configuration. This is consistent with the fact that we see a stronger Raman shift in the 4K sweep than the mK sweep, which was performed after the 4K sweep. 

Knowing that the Raman modes move around stochastically in a manner that may be linked to microscopic details, one may consider a device with multiple individually addressable, non-degenerate Majorana QSLs by patterning different local strain gradients in $\alpha$-RuCl$_3$ on pillars of different sizes. Hysteretic magnetostrictive effects may also be useful for neuromorphic or quantum memory applications, though substantial additional research is essential to corroborate such claims. 


\acknowledgments

The authors would like to acknowledge insightful discussion with Thuc Mai, Rolando Valdes Aguilar, and Arnab Banerjee. 

This research was sponsored by the U. S. Department of Energy, Office of Science, Basic Energy Sciences, Materials Sciences and Engineering Division. Postdoctoral (CEM) research support was provided by the Intelligence Community Postdoctoral Research Fellowship Program at the Oak Ridge National Laboratory, administered by Oak Ridge Institute for Science and Education through an interagency agreement between the U.S. Department of Energy and the Office of the Director of National Intelligence. Portions of this material (JY) are based upon work supported by the U.S. Department of Energy, Office of Science, National Quantum Information Science Research Centers, Quantum Science Center. Exfoliation (KX), variable temperature photoluminescence (BL) and modeling (LL) were performed at the Center for Nanophase Materials Sciences, which is a U.S. Department of Energy Office of Science User Facility. Student (MAF, BEL, YSP) research support was provided by the Department of Defense through the National Defense Science $\&$ Engineering Graduate Fellowship (NDSEG) and by the DOE Science Undergraduate Laboratory Internships (SULI) program.

\bibliography{references}

\end{document}